\documentclass[aps,prd,preprintnumbers,12pt]{revtex4-2}
\usepackage{amsmath, mathrsfs, amssymb,amsfonts,amsthm,graphicx, epsf, dcolumn, yfonts}
\usepackage[hyperfootnotes=true]{hyperref}
\usepackage{tikz}
\usepackage{color}
\usepackage{slashed}
\usepackage{setspace}
\usepackage{cancel}
\usepackage{wasysym}
\usepackage[lofdepth,lotdepth]{subfig}
\usepackage{float}
\usepackage[normalem]{ulem}
\usepackage[symbol]{footmisc}
\pdfoutput=1
\usepackage[utf8]{inputenc}
\usepackage{comment}
\usepackage{ragged2e}
\DeclareCaptionJustification{justified}{\justifying}
 \newcommand{\be}{\begin{equation}}
 \newcommand{\ee}{\end{equation}}
 \newcommand{\bea}{\begin{eqnarray}}
 \newcommand{\eea}{\end{eqnarray}}

\newcommand{\beq}{\begin{equation}}
\newcommand{\eeq}{\end{equation}}

\renewcommand*{\thefootnote}{\fnsymbol{footnote}}

\begin{document}

\thispagestyle{empty}

\preprint{\footnotesize{\texttt{IFT-UAM/CSIC-25-46}}}

\title{Quantum censors: backreaction builds horizons}

\author{Antonia M. Frassino,$^{1,2,3,a}$ Robie A. Hennigar,$^{4,b}$ Juan F. Pedraza$^{5,c}$ and Andrew Svesko$^{6,d}$}
\affiliation{\vspace{1mm}
$^1$SISSA 
\& INFN Sezione di Trieste, 34136 \& 34127 Trieste, Italy\vspace{-1mm}\\
\vspace{-2mm}$^2$Departamento de F\'{i}sica y Matem\'{a}ticas, Universidad de Alcal\'{a},\\ Alcal\'a de Henares, 28805 Madrid, Spain\vspace{-1mm}\\
$^3$
  Institut de Ciències del Cosmos, Universitat de Barcelona, 08028 Barcelona, Spain\vspace{-1mm}\\
\vspace{-2mm}$^4$Centre for Particle Theory, Department of Mathematical Sciences,\\
Durham University, Durham DH1 3LE, UK\vspace{-1mm}\\
$^5$Instituto de F\'isica Te\'orica UAM/CSIC, Cantoblanco, 28049 Madrid, Spain\vspace{-1mm}\\
$^6$Department of Mathematics, King’s College London, Strand, London, WC2R 2LS, UK}

\begin{abstract}\vspace{-1.5mm}
\noindent

\noindent Cosmic censorship posits spacetime singularities remain concealed behind event horizons, preserving the determinism of General Relativity. While quantum gravity is expected to resolve singularities, we argue that cosmic censorship remains necessary whenever spacetime has a reliable semi-classical description. Using holography to construct exact solutions to semi-classical gravity, we show backreaction of quantum matter generates horizons ---quantum censors--- to thwart potential violations of censorship. Along with a quantum Penrose inequality, this provides compelling evidence cosmic censorship is robust, even nonperturbatively, in semi-classical gravity.

\vspace{3.5cm}

\begin{center}
\emph{This essay received  Honorable Mention in the 2025 Essay Competition\\ of the Gravity Research Foundation}
\end{center}
$\,$\\\vspace{-2mm}
\hspace{-8mm}$^a$\verb"afrassin@sissa.it"\\\vspace{-2mm}
\hspace{-8mm}$^b$\verb"robie.a.hennigar@durham.ac.uk"\\\vspace{-2mm}
\hspace{-8mm}$^c$\verb"j.pedraza@csic.es"\\\vspace{-2mm}
\hspace{-8mm}$^d$\verb"andrew.svesko@kcl.ac.uk" 

\end{abstract}

\renewcommand*{\thefootnote}{\arabic{footnote}}
\setcounter{footnote}{0}

\maketitle

\newpage

\setcounter{page}{1}

Fifty years after its inception, Roger Penrose proved General Relativity is incomplete~\cite{Penrose:1964wq}. 
This is because, generically, gravitational collapse leads to singularities, regions in spacetime where Einstein’s field equations no longer apply, and the laws of physics suffer a fatal indeterminacy.
Penrose further postulated that black holes hold a remedy for this fate: singularities developed during gravitational collapse are ``clothed'' by a black hole's event horizon. Colloquially put, there are no \emph{naked} singularities \cite{Penrose:1969pc}. This conjecture, dubbed (weak) cosmic censorship, so far remains unproven. To test its validity, one is therefore left to gather circumstantial evidence for or against  
weak cosmic censorship (WCC).

A classic test of WCC is to try to destroy the horizon of a (near-) extremal black hole by overspinning or overcharging it \cite{Wald:1974hkz,Hubeny:1998ga}. Such attempts have proven impossible in classical general relativity \cite{Sorce:2017dst}. Recently, however, near-extremal black holes were found to be highly susceptible to quantum effects \cite{Iliesiu:2020qvm}, such that, without low energy supersymmetry (as in the real world), exactly extremal black holes do not exist since the classical picture entirely breaks down. This calls into question the classical tests of weak cosmic censorship.

In fact, all known violations of WCC force us to grapple with quantum effects. For example, the final stages of the Gregory-Laflamme instability of black strings \cite{Gregory:1993vy} drive the spacetime geometry to regions where naked curvature singularities appear, and on scales where quantum effects should not be ignored. Similar observations have been made when studying late time evolution of black holes formed under critical collapse \cite{Christodoulou:1984mz}.

A way forward is to aim for singularity resolution. After all, if there are no singularities, there is no need for censorship. As far as we know, resolving spacetime singularities requires a radical departure from classical gravity. Indeed, that general relativity predicts its own breakdown is a herald for a more fundamental description of gravity, namely, quantum gravity. Accounting for quantum gravitational effects, the above violations are rendered moot as the curvature singularity is dissolved \cite{Emparan:2024mbp}.
This is unsurprising as singularity resolution, a traditional benchmark for candidate models of quantum gravity, relies on being in domains where classical spacetime is no longer sensible. In other words, so long as we restrict to regimes where spacetime has a classical description (thus precluding quantum gravitational effects that resolve singularities) cosmic censorship is necessary.

There are quantum effects that do not resolve singularities.  
Indeed,  in semi-classical gravity, where spacetime responds classically to quantum matter,
classical singularity theorems are superseded by
\emph{quantum} singularity theorems \cite{Wall:2010jtc,Fewster:2021mmz}. That is, 
solutions to semi-classical Einstein's equations sourced by quantum stress-energy $\langle T^{\text{mat}}_{\mu\nu}\rangle$,
\beq R_{\mu\nu}-\frac{1}{2}g_{\mu\nu}R+\Lambda g_{\mu\nu}=8\pi G_{\text{N}}\langle T^{\text{mat}}_{\mu\nu}\rangle\;,\label{eq:semieineom}\eeq
generically develop singularities. Quantum effects featured in semi-classical gravity are distinct from bona fide quantum gravitational effects. Thus, when spacetime has a reliable semi-classical geometric description, the setting we macroscopic observers regularly experience, singularities are inevitable. Notably, however, Penrose's classical censorship conjecture is now violated due to the backreaction of these quantum matter effects, even when backreaction results in perturbatively small corrections to the classical geometry \cite{Bousso:2019var,Bousso:2019bkg}.

To be deterministic, semi-classical gravity therefore must be accompanied by an appropriate generalization of cosmic censorship, viz. \emph{quantum} cosmic censorship.

\vspace{2mm}

\noindent \textbf{Toward quantum cosmic censorship.} A naive notion of (weak) quantum cosmic censorship is easy to state: there are no naked \textit{quantum} singularities.  Should a singularity develop, none of its effects propagate to a distant observer. The singularities of collapse should therefore be hidden inside \textit{quantum} black holes --- the black hole solutions of the semi-classical Einstein equations. To be more precise, first recall a formal statement of classical WCC:

\vspace{1mm} 

\noindent \emph{Classical Weak Cosmic Censorship:} Let $\Sigma$ be a codimension-1 manifold that is topologically a connected sum $\mathbb{R}^{D-1}$ (for $D\geq3$) and a compact manifold with induced metric $h_{\mu\nu}$ and extrinsic curvature $K_{\mu\nu}$. Let $(h_{\mu\nu},K_{\mu\nu},\psi)$ be non-singular asymptotically \emph{flat} initial data on $\Sigma$ for a solution to the \emph{classical} Einstein's equations with ``suitable'' matter (here $\psi$ denotes initial data for said matter). Then, ``generically'', the maximal Cauchy evolution of this data is an asymptotically \emph{flat} $D$-dimensional spacetime $(M,g_{\mu\nu})$ at future null infinity with complete conformal boundary $\mathcal{J}^{+}$. \label{conj:WCC}

\vspace{1mm}

The asymptotic flatness condition with complete $\mathcal{J}^{+}$ captures the notion that asymptotic observers are ignorant to the singularity. Note the conjecture can be appropriately extended to asymptotic anti-de Sitter spacetimes, which have a timelike conformal boundary.  Further, as stated, the conjecture is imprecise for two reasons: what constitutes (i) ``suitable'' matter, and (ii) ``generic'' evolution. For the former, it is expected the Einstein-matter field equations have a well posed initial value formulation, and the classical matter stress-tensor obeys appropriate energy conditions, e.g., the dominant or null energy conditions. Such assumptions are also crucial for the validity of the classical singularity theorems. Condition (ii) is to deal with technicalities, superficially-tuned examples designed to violate WCC.

A very naive formulation of weak quantum cosmic censorship would then be to simply replace ``\emph{classical}'' in the conjecture with \emph{semi-classical}. The resulting semi-classical conjecture is undoubtedly less precise than its classical counterpart. The matter is now treated quantum mechanically, and it is not immediately obvious what constitutes ``suitable'' quantum matter. Presumably, the classical energy conditions, which are known to be violated by quantum matter (and thus violate the assumptions underlying classical singularity theorems), should be replaced by ``quantum energy conditions''. 
More problematic is how to understand ``generic'' evolution of initial data. This is because we have very little understanding of the dynamics of the semi-classical Einstein equations (\ref{eq:semieineom}). Indeed, very few exact black hole solutions to (\ref{eq:semieineom}) are known -- a point we will return to shortly. Thus, directly formulating weak quantum cosmic censorship is far more difficult than classical WCC.

An indirect approach toward formulating weak quantum cosmic censorship, is arguably more tractable. One such approach relies on our intuition from classical gravity. 
Predating the classical tests of WCC, Penrose tried to formally uncover counterexamples to WCC, leading him to propose a spacetime inequality quantifying the mass $M_{\text{ADM}}$ of said
spacetime in terms of the horizon area $A$ of the black holes it contains \cite{Penrose:1973um}, 
\beq G_{4}M_{\text{ADM}}\geq\sqrt{\frac{A[\sigma]}{16\pi}}\;,
\label{eq:PenroseinqOG}\eeq
for marginally trapped surface $\sigma$. 
The conjectured Penrose inequality (PI) may be generalized to higher-dimensions $D\geq4$ and asymptotically anti-de Sitter (AdS) spacetimes.

The classical Penrose inequality rests on two assumptions: weak cosmic censorship, and that regular initial data settles to a stationary black hole. Thus, the PI serves as a necessary condition to WCC; violations of the PI often imply violations of weak cosmic censorship. Although the PI is a weaker statement compared to cosmic censorship, it is at least a quantitative relation which can and has been extensively explored. Notably, there are no classical counterexamples to (\ref{eq:PenroseinqOG}), providing strong circumstantial evidence that WCC holds.

The classical inequality, though, is violated by semi-classical quantum effects. For example, negative energy due to quantum fields (in the Boulware vacuum) near a static AdS black hole horizon negatively contributes to the mass such that (\ref{eq:PenroseinqOG}) is violated. Naturally, this motivates whether a semi-classical generalization exists. Such a quantum Penrose inequality (QPI) was proposed for $D\geq4$ asymptotically AdS spacetimes with length scale $\ell_{D}$ \cite{Bousso:2019bkg}
\beq
\begin{split} 
\hspace{-4mm}\frac{16\pi G_{D}M_{\text{AMD}}}{(D-2)\Omega_{D-2}}&\geq \left(\frac{4G_{D}S_{\text{gen}}}{\Omega_{D-2}}\right)^{\hspace{-1mm}\frac{D-3}{D-2}}\hspace{-2mm}+\ell^{-2}_{D}\left(\frac{4G_{D}S_{\text{gen}}}{\Omega_{D-2}}\right)^{\hspace{-1mm}\frac{D-1}{D-2}}\;.
\end{split}
\label{eq:qAdSPI}\eeq
Notably, the area $A$ has been replaced by generalized entropy $S_{\text{gen}}$, the sum of the gravitational entropy of a codimension-2 Cauchy splitting surface and the entropy of the quantum matter confined to one side of the Cauchy splitting surface.

The QPI (\ref{eq:qAdSPI}) holds for weak backreaction \cite{Bousso:2019bkg}. To fully assess its validity, it is crucial to test it also when backreaction is strong.
 At a minimum, this requires self-consistent black hole solutions to the semi-classical equations (\ref{eq:semieineom}) in $D\geq4$, a difficult and open problem.

\vspace{2mm}

\noindent \textbf{Exact quantum black holes.} Uncovering exact solutions to the semi-classical Einstein equations (\ref{eq:semieineom}) is difficult in practice as it requires simultaneously solving a coupled system of geometry and quantum correlators. Loosely, here is how the algorithm would go: \textbf{(i)} Compute the renormalized quantum matter stress-tensor $\langle T_{\mu\nu}^{\text{mat}}\rangle$ in a fixed classical background. \textbf{(ii)} Source the Einstein equations with the quantum stress-tensor and iteratively solve the resulting non-linear partial differential equations to uncover the new geometry. Step (i) is already computationally very challenging, but step (ii) remains an infamously open problem.

To make progress, one approach, and the one we take here, is to appeal to gravitational holography. More specifically, `double' or `braneworld holography' \cite{deHaro:2000wj}, where
a $(D-1)$-dimensional membrane (`braneworld') is embedded inside an asymptotically AdS$_{D}$ spacetime that
has a dual description in terms a non-gravitating conformal field theory living on its boundary. 
More precisely, in double holography, a portion of the `bulk' AdS spacetime, including its boundary, is removed by an end-of-the-world (ETW) brane. 
Crucially, the brane geometry has intrinsic dynamics characterized by an induced semi-classical theory of higher-derivative gravity.
The higher-derivative corrections to Einstein gravity arise because the ETW brane acts as an ultraviolet (UV) cutoff for the  CFT, identical to holographic renormalization.  
Heuristically, the CFT living on a portion of the bulk AdS boundary has been pushed inward to the brane. From the brane perspective, the higher-derivative corrections incorporate backreaction effects due to this CFT living on the brane.

The practical advantage of double holography is that the induced semi-classical theory
is equivalent to $D$-dimensional General Relativity coupled to a brane obeying Israel junction conditions. Thus, spacetimes solving the classical bulk field equations with brane boundary conditions automatically correspond to exact solutions to the semi-classical brane equations of motion.
In particular, classical $\text{AdS}_{D}$ black holes that localize on the brane exactly map to quantum black holes in $(D-1)$-dimensions, including all orders of quantum backreaction, and, in principle, for any quantum matter state \cite{Emparan:2002px}.

Exact quantum black holes can be analytically found using three-dimensional braneworlds \cite{Emparan:2002px,Emparan:2020znc,Panella:2024sor}. For example, the static, neutral `quantum BTZ' (qBTZ) black hole is \cite{Emparan:2020znc} 
\beq
\begin{split}
&{\rm d}s^2=-f(r){\rm d}t^2+\frac{{\rm d}r^2}{f(r)}+r^2{\rm d}\phi^2\,,\qquad f(r)=\frac{r^2}{\ell_3^2}-8\mathcal{G}_3M-\frac{\ell F(M)}{r}\,.
\end{split}
\label{eq:qBTZmain}\eeq
Here $\mathcal{G}_3$ is the `renormalized' 3D Newton's constant, $F(M)$ is a positive function of the mass, and $\ell_{3}$ is the $\text{AdS}_{3}$ length scale. The metric (\ref{eq:qBTZmain}) is an exact black hole solution to the induced semi-classical theory, where the non-negative parameter $\ell\sim c\, L_P$ controls the strength of backreaction from a large-$c$ $\text{CFT}_{3}$. Notably, the scale $\ell$ is macroscopic, i.e., much larger than the Planck length $L_{P}\sim \hbar\, G_3$. In contrast with perturbative constructions \cite{Casals:2016ioo}, where the $\ell F(M)/r$ enters at order $L_{P}$ and would thus be inconsistent to neglect quantum gravitational effects, the solution (\ref{eq:qBTZmain}) is fully within the validity of semi-classical gravity.

\begin{figure}[t!]
\includegraphics[clip, trim=5mm 60mm 3mm 65mm, width=0.75\textwidth]{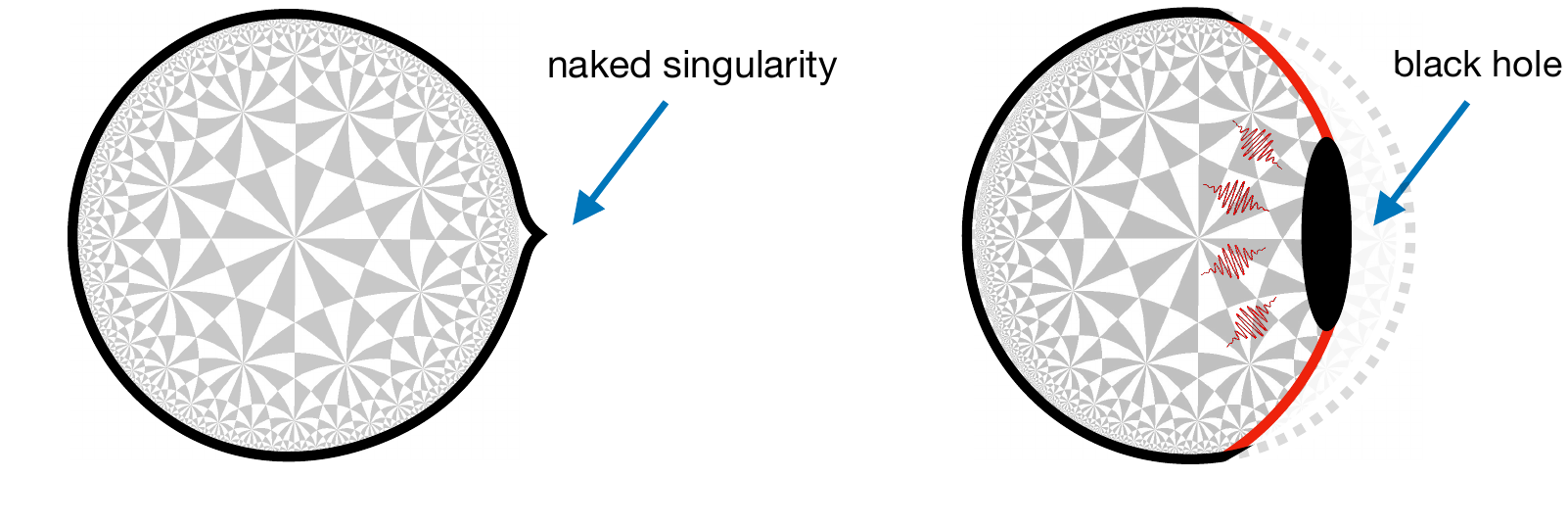}
\put(-299,-2){$\hbar=0$}
\put(-97,-2){$\hbar\neq0$}
\caption{\label{fig:qBH} Left: AdS/CFT setup with a boundary naked singularity. Right: a (quantum) black hole localized on an ETW brane (red line), where a cutoff CFT couples to gravity.}
\end{figure}

The semi-classical geometry represents a family of black holes with two distinct origins: quantum-corrected black holes, or quantum-corrected conical defects. That is, recall for masses $-1/(8G_{3})<M<0$ the classical BTZ solution describes a defect with a naked conical singularity, introducing a gap in the continuous mass spectrum. 
Quantum backreaction, then, induces a horizon such that the conical singularities become censored and the mass gap is removed. 
This intuitively captures the spirit of weak quantum cosmic censorship (Figure \ref{fig:qBH}). 
It is also reassuring that the (rotating) quantum BTZ black hole cannot be overspun to shed its horizon \cite{Frassino:2024fin}, despite exhibiting superradiant instabilities \cite{Cartwright:2025fay}.

\vspace{2mm}

\noindent \textbf{Exact quantum Penrose inequality.} Equipped with exact quantum AdS$_{3}$ black holes, the quantum Penrose inequality (\ref{eq:qAdSPI}) can be explicitly tested beyond regimes of perturbative backreaction. A naive application of (\ref{eq:qAdSPI}) with $D=3$
reveals violations appear beyond the perturbative regime \cite{Frassino:2024bjg}. This is unsurprising since 
 the classical AdS$_{3}$ Penrose inequality is already more subtle than for AdS$_{D\geq4}$.
 Specifically, unlike in $D\geq4$, AdS$_{3}$ black holes formed under collapse cannot have arbitrarily small mass due to the mass gap. This imposes an additional criterion on the AdS$_{3}$ initial data to be used in a heuristic derivation of the classical PI: initial data mass cannot go below the mass gap. 

Instead, we find the following quantum inequality \cite{Frassino:2024bjg}
\beq 8\mathcal{G}_{3}M_{\text{AMD}}\geq\ell_{3}^{-2}\left(\frac{4\mathcal{G}_{3}S_{\text{gen}}}{2\pi}\right)^{\hspace{-1mm}2}\;,\label{eq:AdS3qPeninq}\eeq
exactly holds for all quantum black holes for $\ell\geq 0$. 
For $\ell\neq0$, inequality (\ref{eq:AdS3qPeninq}) strictly holds for black holes with positive temperature and entropy. Notably, the inequality is not saturated for the static qBTZ black hole, except when the black hole shrinks to arbitrarily small size. 

Our analysis links the inability to saturate Penrose inequalities to the existence of a gap in the mass spectrum. Indeed, our classical AdS$_{3}$ PI (when $\ell\to0$) is not saturated for the static BTZ black hole, in contrast with the $D\geq4$ inequality. 
Meanwhile, the qBTZ solution has a continuous mass spectrum. Thence, the quantum effects that shrink the mass gap are also responsible for the saturation of the quantum PI (\ref{eq:AdS3qPeninq}) for small black holes.

That the inequality (\ref{eq:AdS3qPeninq}) holds for all known AdS$_{3}$ quantum black holes, for small and large backreaction, strongly indicates weak quantum cosmic censorship is robust, valid beyond perturbative semi-classical gravity. Backreaction builds horizons that serve as quantum censors, safeguarding the integrity of spacetime predictability.


\noindent \emph{Acknowledgements.} 
We are grateful to José Barbón, Casey Cartwright, Roberto Emparan, Umut G\"ursoy, Clifford Johnson, Eleni Kontou, David Kubiz{\v n}{\' a}k, José Navarro-Salas, Emanuele Panella, Jorge Rocha, Marija Toma\v{s}evi\'{c} and George Zahariade for illuminating discussions and correspondence.  AMF acknowledges the support from EU Horizon 2020 Research and Innovation Programme under the Marie Sklodowska-Curie Grant Agreement No 101149470.
 JFP is supported by the `Atracci\'on de Talento' program grant 2020-T1/TIC-20495 and by the Spanish Research Agency through the grants CEX2020-001007-S and PID2021-123017NB-I00, funded by MCIN/AEI/10.13039/501100011033 and by ERDF `A way of making Europe.' AS is supported by STFC grant ST/X000753/1.

\bibliography{qineqrefs}

\end{document}